
\input harvmac
\parskip 6pt
\def\sqr#1#2{{\vcenter{\hrule height.#2pt \hbox{\vrule width.#2pt
height#1pt \kern#1pt \vrule width.#2pt} \hrule height.#2pt}}}

\def \darrow#1{\raise1.5ex\hbox{$\rightarrow$}\mkern-16.5mu #1}
\def\darr#1{\raise1.5ex\hbox{$\leftrightarrow$}\mkern-16.5mu #1}

\def\rpartial{\darrow{\partial}}
\def\rdelta{\darrow{\delta}}
\def\phi{\varphi}

\def\ie{{\it i.e.}\ }
\def\eg{{\it e.g.}\ }

\def\viz{{\it viz.}\ }

\def\aka{{\it a.k.a.}\ }

\noblackbox
\pageno=0\nopagenumbers\tolerance=10000\hfuzz=5pt
\baselineskip=12pt plus2pt minus2pt
\newskip\footskip\footskip=10pt 
\rightline{hep-th/0008123}
\rightline{SHEP 00-09}
\vskip 10pt
\centerline {\bf  Exact Scheme Independence }
\vskip 30pt
\centerline {Jos\'e I. Latorre$^{a}$
and Tim R. Morris$^{b}$}
\vskip 20pt

\centerline{${}^{(a)}${\it
Departament d'Estructura i Constituents de la Mat\`eria,}}
\centerline{\it Universitat de Barcelona}
\centerline{and}
\centerline{{\it I.F.A.E.,}}
\centerline {\it Diagonal 647, E-08028 Barcelona, Spain}
\vskip 10pt
\centerline{${}^{(b)}$\it 
Department of Physics, University of Southampton,}
\centerline{\it Highfield, Southampton SO17 1BJ, UK}

\vskip 40pt

{\centerline{\bf Abstract}
\medskip\narrower
\baselineskip=10pt

Scheme independence of exact renormalization group
equations, including independence of the choice of cutoff function,
 is shown to follow from general field redefinitions, which
remains an inherent
redundancy in quantum field theories. Renormalization group equations
and their solutions are amenable to a simple formulation which is
manifestly covariant under such a symmetry group. Notably, the
kernel of the exact equations which controls the integration of modes
acts as a field connection along the flow.
\smallskip}
\vskip 48pt

\vskip 64pt
\line{August 2000\hfill}

\vfill\eject \footline={\hss\tenrm\folio\hss}
\nref\kogwil{K. G. Wilson and J. Kogut, Phys. Rep. 12C (1974) 75.}
\nref\wegho{F. J. Wegner and A. Houghton, Phys. Rev. A8 (1972) 401. }
\newsec{Introduction}
Quantum field theory provides a redundant tool to describe
physics. A particular manifestation of this fact corresponds
to the freedom to define a renormalization
scheme in order for perturbative as well as non-perturbative computations 
to make explicit contact with observables. 
Independence of  observables on different choices of scheme
may be obscure if not impossible 
to check in practice, but it is a continuum property
inherent to quantum field theory.
In the framework of Wilson's Renormalization Group (RG) and in particular
the exact RG \kogwil, this property is especially deeply embedded and scheme
independence is an issue that it is 
essential to address.
Thus, a particular form of the exact RG receives its very definition
in part by specifying a cutoff function. Beyond this, it is
clear that one also has considerable further freedom
in formulating flow equations
\ref\red{F.J. Wegner, J. Phys. C7 (1974) 2098.}
\nref\alg{T. R. Morris, 
in {\it The Exact Renormalization Group}, Eds Krasnitz {\it et al},
World Sci (1999) 1, and hep-th/9810104.}--\nref\ymi{
    T. R. Morris,
    Nucl. Phys. B573 (2000) 97, hep-th/9910058.}\ref\ymii{
    T. R. Morris,
    hep-th/0006064.}.
This extra freedom in choosing a scheme
can be put to great advantage \alg--\ymii. We comment
further below and in the conclusions.

The purpose of this note is to show that many changes of scheme
in an exact RG equation can be implemented as field redefinitions.
It is then clear that, within 
such a class of RGs, although the cutoff
function and/or the form of the flow equation
will vary, no universal properties will be modified; no
observable prediction for the continuum quantum field theory
will change. 
We shall proceed in three steps. First, we shall review what
is already known and reanalyse 
Polchinski's exact RG equation in order to provide a simple
understanding of its parts.  Second, we shall perform  
a general field redefinition to show that
its effect is to recast the original equation into 
one written in a different form. A change of cutoff function
is seen to be just a particular case of the freedom to
perform field redefinitions. Third, we take a
broader view and show that the exact
RG equation operates as a covariance statement. The final form
and understanding of the exact RG equation will be very simple.

Note that there are very many ways to derive the exact RGs 
currently in use \kogwil\wegho\alg\ref\wein{S. Weinberg, Erice lectures,
Subnucl. Phys. (1976) 1.}\nref\nici{J. F. Nicoll 
and T. S. Chang, Phys. Lett. 62A (1977) 287.}\nref\pol{
J. Polchinski, Nucl. Phys. B231 (1984) 269.}
\nref\wet{C. Wetterich, 
Phys. Lett. B301 (1993) 90.}--\nref\bonini{Bonini {\it et al},
Nucl. Phys. B409 (1993) 441, hep-th/9301114.}\ref\erg{T. R. 
Morris, Int. J. Mod. Phys. A9 (1994) 2411, hep-ph/9308265.}.
 Almost all of these 
derivations yield the same exact RG however, in some cases
after some small transformations. (The underlying reason is that 
all such derivations start effectively by making the 
simplest choice of placing the effective
cutoff $\Lambda$ only in the bilinear kinetic terms.)
Thus the Wegner-Houghton RG \wegho\ is the sharp cutoff limit 
\erg--\nref\truncam{
T. R. Morris, Phys. Lett. B334 (1994) 355, hep-th/9405190  ;\ 
Nucl. Phys. B458[FS] (1996) 477, hep-th/9508017.}
\ref\sumi{J.-I. Sumi {\it et al}, 
hep-th/0002231.} of Polchinski's RG \pol.
Polchinski's RG is transformed into
Wilson's exact RG \kogwil, by the momentum dependent change
of variables $\phi\mapsto\phi\sqrt{K}$ that eliminates 
Polchinski's cutoff function $K$ from the propagator \ref\deriv{T. R. Morris,
Phys. Lett. B329 (1994) 241, hep-ph/9403340.}.\foot{The physical
meaning of the equivalence is not so clear to us
however.} Finally the exact RG for the Legendre effective action,
or effective average action \nici\wet--\erg,
is the Legendre transform of Polchinski's equation \erg\ref\YKIS{T. R. Morris,
in {\it Yukawa International Seminar '97}, 
Prog. Theor. Phys. Suppl. 131 (1998) 395, hep-th/9802039.}.  

There have also been a number of studies of the dependence on the different
forms of exact RG and on the form of the cutoff function within certain
approximation schemes \ref\jose{
R. D. Ball, P. E. Haagensen, J. I. Latorre and E. Moreno,
Phys. Lett. B347 (1995) 80, hep-th/9411122.}
\ref\cdepend{
D. F. Litim, Phys. Lett. B393 (1997) 103, hep-th/9609040;\hfill\break
J. Comellas, Nucl. Phys. B509 (1998) 662, hep-th/9705129;\hfill\break
R. Neves, Y. Kubyshin and  R. Potting,
in {\it The Exact Renormalization Group}, Eds Krasnitz {\it et al},
World Sci (1999), hep-th/9811151;\hfill\break
T. R. Morris and J. F. Tighe, JHEP 08 (1999) 7, hep-th/9906166 ;\hfill\break
D. F. Litim, Phys. Lett. B486 (2000) 92, hep-th/0005245.}

The freedom in formulating exact RGs is more than this, as was originally
pointed out by Wegner \red. He noted
that in principle a large class of exact RGs could
be defined, based on three parts: the integrating out transformation,
\viz ``elimination of variables'' (for some cases), the rescaling
(or ``dilatation'', which may be incorporated by a change to 
scaling-dimensionless variables using $\Lambda$),
and a possible infinitessimal field redefinition
\eqn\redef{\phi_x\mapsto \phi_x+\delta\Lambda\,\Psi_x}
under the small reduction $\Lambda\mapsto\Lambda-\delta\Lambda$.
In fact, importantly,  the first two parts may also be cast as field
redefinitions. In this way, all the exact RGs correspond precisely to
certain reparametrizations of the partition function \alg.

We show this in sec. 2 by showing that rescaling is equivalent to a field
redefinition. The fact that an integrating out transformation
is also equivalent to a field redefinition then follows from
differing derivations of the same exact RG \erg\alg, and
the equivalences between exact RGs recalled above \wegho--\alg.

In sec. 2 we concentrate on the Polchinski equation and establish that
it implies precisely a reparametrization of the partition function
under a scale dependent field redefinition.  In sec. 3 we consider
general field redefinitions and establish the general form and
transformation properties of the resulting exact RGs. As an example we
show that change of cutoff function in Polchinski's exact RG simply
corresponds to one of these field redefinitions. Appendix A contains
some details for this computation and sketches a general method for
finding such field redefinitions.  In sec. 4 we explain how the
transformations work when the $t$ dependence is regarded as only
inherited through the action. We recover the example of sec. 3 from
this point of view.  In sec. 5, we develop further the consequences of
the resulting structures and show that they may all be simply
understood in terms of a {\sl connection} relating different
scales. Finally in sec. 6 we draw our conclusions.

\newsec{Revisiting Polchinski}

To make the discussion concrete, consider Polchinski's form of the exact
RG \pol\ written for the total effective action $S$, 
which we need in its renormalized scaled version \jose:\foot{The unscaled
version was discussed in this light in ref. \alg.}
\eqn\polchinski{
\partial_tS+d\int_p\!\phi_p{\delta S\over\delta\phi_p}
+\int_p\!\phi_p\,
  p^\mu{\partial\over\partial p^\mu}{\delta S\over\delta\phi_p}
= \int_p c'(p^2)
     \left( {\delta S\over \delta \phi_p}{\delta S\over \delta \phi_{-p}}
     - {\delta^2 S\over \delta \phi_p \delta \phi_{-p}}
  -2{p^2\over c(p^2)}\phi_p {\delta S\over \delta \phi_p}\right) \ ,
}
where $t\equiv \ln {\mu\over\Lambda}$, $\mu$ being some fixed 
physical scale
and $\Lambda$ the Wilsonian cutoff which is taken towards 0,
$S[\phi,t]$ is a functional of $\phi$ and a function of $t$.
On the l.h.s., the second term arises from rescaling the field,
$d$ being the full scaling dimension of the field, and the third
term arises from rescaling the momenta (which is often rewritten in
terms of a derivative that does not act on momentum-conserving
$\delta$ functions \wegho\ or as $\Delta_\partial$, the space-time
derivative counting operator \deriv).
The scheme is defined through the choice of 
the cutoff function, which after scaling is just $c(p^2)$
(and thus $c'(p^2)$ means $\partial c(p^2)/\partial p^2$).
It is less obvious at this level that the form
of the r.h.s. is also a choice of scheme. 

Actually, \polchinski\ does not arise from rescaling Polchinski's 
original formulation \pol, unless we set $d$ to the engineering dimension
of the field (\ie $d=D/2-1$ for a scalar field in spacetime dimension
$D$) whereas we mean to incorporate also the anomalous dimension
($d(t)=D/2-1+\gamma(t)/2$ for a scalar field) \jose\sumi.
Nevertheless, \polchinski\ is also a {\sl perfectly valid} exact RG 
\jose\alg\ which can be justified on the grounds that $\gamma$ 
can indeed be chosen 
as usual to constrain the normalisation of the kinetic term \jose, that 
\polchinski\ still leaves the partition function invariant \jose\sumi\
(as we will confirm below) and most importantly that \polchinski\ 
still corresponds to integrating out. (This latter follows on quite 
general grounds, 
because the partition function is left invariant as $\Lambda$ decreases,
but since all quantum corrections are UV regularised by $\Lambda$,
\ie limited above by $\Lambda$, the lost modes are being incorporated
in $S$ \ymi.) 
This is then one small but important example of the freedom allowed in
the definition of the exact RGs.

The exact RG equation \polchinski\ 
can also be written in a more enlightening form
\eqn\wegner{
  \partial_t\, {\rm e}^{-S}=
  \partial_\alpha\left( \Psi^\alpha {\rm e}^{-S}\right)\ ,
}
where
\eqn\psidef{
\Psi^\alpha[\phi,t]= \left( D-d(t)\right)
\phi_p+p^\mu{\partial\phi_p\over\partial p^\mu}\,+\,
c'(p^2)\left(- {\delta \over \delta \phi_{-p}}
     -2{p^2\over c(p^2)} \phi_p\right)\ ,
}
discarding vacuum energy terms,\foot{corresponding
to the Jacobian for scaling, and a term 
discarded in \pol.}
and we have introduced the compact notation
\eqn\notation{
  \partial_\alpha \equiv {\delta\over \delta \phi_p}\ ,
}
(integration over $p$ being implied by the summation convention.
The sub- and super-index $\alpha$ can also stand for
any further quantum number associated with the field $\phi$.)

Note that \wegner\ immediately implies that the partition function
\eqn\ptnfn{{\cal Z}= \int {\cal D}\phi \ {\rm e}^{-S}} 
is left invariant by \polchinski, \ie $\partial_t\,{\cal Z}=0$, as
claimed.  Furthermore in this form, the r.h.s. of the exact RG
equation corresponds to the integrand of a field redefinition \alg, as
we now explain in detail.

For an infinitessimal field redefinition
$\tilde\phi^\alpha=\phi^\alpha-\theta^\alpha[\phi,t]$ (optionally scale
dependent) the Jacobian is $\left|{\delta\phi/ \delta\tilde\phi}
\right|= 1+ \partial_\alpha \theta^\alpha$.  Together with the term
from the action itself, the changes in ${\cal Z}$ combine to produce
the identity \eqn\fieldredefid{ \int {\cal D}\phi\
\partial_\alpha\left(\theta^\alpha {\rm e}^{-S} \right)=0\ .  } 
Thus from \wegner, an infinitessimal step in $t$ to $t+\delta t$,
corresponds to the field redefinition
\eqn\psiredef{\tilde\phi^\alpha=\phi^\alpha-\delta
t\,\Psi^\alpha[\phi,t]\ .}

This is in effect the way Polchinski's equation was originally
deduced and brings separate meanings to
both left and right hand sides of the exact RG equation. If integrated
\eqn\newwegner{
 0=\partial_t \int {\cal D} \phi \ {\rm e}^{-S}
= \int {\cal D}\phi \  \partial_\alpha\left(\Psi^\alpha {\rm e}^{-S}
  \right)=0\ ,
}
the l.h.s. vanishes because low-energy  observables do not depend 
on the cut-off, whereas the r.h.s. is zero because
it is the total derivative term emerging from
a field redefinition as  in \fieldredefid. 

Thus we may regard 
the underlying structure of an exact RG equation
as a trade-off.
A change of cut-off is compensated by a field redefinition.

\newsec{General field redefinitions}

We are near to establishing the transformation
properties of all the elements of \wegner\ under field redefinitions.
We shall now complete this analysis. 

For the sake of precision let us start by recalling that the dependences of
the kernel of 
the exact RG equation are
\eqn\dependpsi{
 \Psi=\Psi^\alpha[\phi,t]\ ,
}
and that we want to analyse its behaviour under an infinitessimal field
redefinition
\eqn\filedredefdep{
 \phi^\alpha \longrightarrow \tilde\phi^\alpha=\phi^\alpha 
-\theta^\alpha[\phi,t]
 \ .
}
On the one hand, we know 
that the integrand of the path integral $\exp{(-S)}$
transforms as a density on absorbing the Jacobian of the field
reparametrization:
\eqn\density{ {\rm
e}^{-\tilde S} = \left|{\delta\phi\over \delta\tilde\phi}\right| {\rm
e}^{-S}\ .  } 
Similarly we demand that  the r.h.s. of \wegner\  also
 transforms as a density \ie
\eqn\rhstrans{
 \tilde\partial_\alpha \left(\tilde\Psi^\alpha {\rm e}^{-\tilde 
 S[\tilde\phi]}\right) =
 \left| {\delta \phi\over \delta\tilde\phi}\right| \partial_\alpha
 \left(\Psi^\alpha {\rm e}^{-S[\phi]}\right) \ .
}
This follows 
given \density, if we require that $\Psi^\alpha$
transforms as a vector, $\tilde\Psi^\alpha={\delta \tilde\phi^\alpha\over
 \delta \phi^\beta} \Psi^\beta$. On the other hand the l.h.s.
transforms under a field redefinition as
\eqn\manipul{\eqalign{
   \left.\partial_t\right|_{\tilde\phi}
    {\rm e}^{-\tilde S[\tilde \phi]}&=
   \left(\left.\partial_t\right|_\phi
       + \left.\partial_t \phi^\gamma
        \right|_{\tilde\phi} \partial_\gamma\right)
       \left( \left| {\delta \phi\over \delta\tilde\phi}\right|
        {\rm e}^{-S}\right)\cr
&=\left| {\delta \phi\over \delta\tilde\phi}\right|
\left.\partial_t\right|_{\phi}{\rm e}^{-S}
+{\rm e}^{-S}\left.\partial_t\right|_{\phi}
\left| {\delta \phi\over \delta\tilde\phi}\right|
+\left.\partial_t \phi^\gamma
        \right|_{\tilde\phi} \partial_\gamma
       \left( \left| {\delta \phi\over \delta\tilde\phi}\right|
        {\rm e}^{-S}\right)\ .
}}
Substituting \wegner\ and \rhstrans\ 
for the first term, and 
expanding the other two to first order in $\theta^\alpha$, 
\eqn\finaltilde{
  \eqalign{
    \left.\partial_t\right|_{\tilde\phi}
    {\rm e}^{-\tilde S}
    =&
    \tilde\partial_\alpha\left(
     \left(\tilde \Psi^\alpha + \partial_t \theta^\alpha\right)
      {\rm e}^{-\tilde S}
    \right)\cr
    \equiv & \tilde\partial_\alpha
     \left(\hat \Psi^{\alpha} {\rm e}^{-\tilde S}
     \right)\cr
  }
}
Thus, a field redefinition preserves the covariant form of the equation
except for a shift in the functional form of the kernel.
We may describe the change of $\Psi$ just in terms of the original
fields $\phi$. Using the vector transformation law we get
\eqn\psitranf{
  \eqalign{
   \delta \Psi^\alpha[\phi,t]
       &\equiv \hat\Psi^{\alpha}[\phi,t]-\Psi^\alpha[\phi,t]\cr
       &= \partial_t \theta^\alpha
        +\theta^\beta\partial_\beta\Psi^\alpha-\Psi^\beta\partial_\beta
         \theta^\alpha\ .\cr 
  }
}
This can  be further framed as a standard vector field transformation
plus the explicit $t$ derivative,
\eqn\vectortransf{
\delta\Psi^\alpha \rpartial_\alpha =
 \partial_t \theta^\alpha  \rpartial_\alpha+
  \left[ \theta^\beta\rpartial_\beta,\Psi^\alpha\rpartial_\alpha\right]
\ ,}
where we have introduced the notation $\rpartial_\alpha$ to emphasise
that here the $\partial_\alpha$ is understood to act on all possible
terms appearing on its right.\foot{up to eventually a
functional on the far r.h.s
\eg ${\rm e}^{-S}$, which is here only implicit}
A field redefinition modifies the kernel of the exact RG equation
in this  way. (Ref. \ref\wetfr{C. Wetterich, Z. Phys. C72 (1996) 139, hep-ph/9604227.}\ gives expressions for
the transformation of the flow equation of the Legendre effective action, 
a.k.a. effective  average action, under field redefinitions.)

Actually from \fieldredefid\ and \wegner, 
it is more appropriate for us to 
put the differential on the left. Let us define the following shorthand
\eqn\gadf{\theta\equiv \rpartial_\alpha\theta^\alpha\qquad,\qquad
A_t\equiv \rpartial_\alpha\Psi^\alpha\qquad,\qquad
\delta A_t\equiv \rpartial_\alpha\delta \Psi^\alpha\qquad{\rm and}\qquad
D_t\equiv\partial_t-A_t\ .}
In this notation the exact RG \wegner\ simply reads
\eqn\simplerg{
   D_t {\,\rm e}^{-S}=0\ .
}
By \fieldredefid, the change in the action under a field redefinition 
is just
\eqn\gas{\delta {\,\rm e}^{-S}= \theta {\,\rm e}^{-S}}
and, either by integrating by parts the derivative in \vectortransf,
or directly from \psitranf,
\eqn\gaa{\delta A_t=\partial_t\theta-[A_t,\theta]\ 
\equiv [D_t,\theta]\ .}
Now it is easy to see that our transformation for $\Psi^\alpha$
does indeed ensure that the exact RG transforms covariantly:
\eqn\gaerg{\eqalign{\delta\left(D_t {\,\rm e}^{-S}\right)
&=-\delta A_t{\,\rm e}^{-S}+D_t\,\delta{\,\rm e}^{-S}\cr
&=-[D_t,\theta]{\,\rm e}^{-S}+ D_t\, \theta{\,\rm e}^{-S}\cr
&=\theta\left(D_t {\,\rm e}^{-S}\right)\ .}}

Let us illustrate the above discussion with the particular example of 
changing the cutoff in Polchinski's equation. For the moment we set
$\gamma=0$. The $\gamma\ne0$ case is a little more involved and is
solved in appendix A. First we note from \psidef,
that acting on ${\rm e}^{-S}$ we may equivalently write
\eqn\polpsi{
  A_t\equiv \int_q {\rdelta\over \delta\phi_q}\left(
(D-d)\,\phi_q+q^\mu{\partial\over\partial q^\mu}\phi_q -c'(q^2)
 {\rdelta\over \delta\phi_{-q}}
 - 2 {c'(q^2)\over c(q^2)} q^2\phi_q \right)
  \ .
}
We consider the infinitessimal field redefinition
\eqn\sumif{
\theta^\alpha={1\over2}{\delta c(q^2)\over q^2}
{\delta S\over\delta\phi_{-q}}
-{\delta c(q^2)\over c(q^2)}\phi_q\ ,}
which in \gas, may equivalently be written
\eqn\polf{\theta\equiv-{1\over 2} \int_q {\delta c(q^2)\over q^2}
{\rdelta\over \delta\phi_q}\left( {\rdelta\over \delta\phi_{-q}}
 + 2 {q^2\over c(q^2)} \phi_q\right)\ .}
This can be derived by transforming the results of refs. 
\sumi--\ref\gol{E. K. Riedel, G. R. Golner and K. E. Newman, 
Ann. Phys. 161 (1985) 178.}\ref\bervil{C. Bagnuls and  
C. Bervillier, hep-th/0002034.}\ to the full
action, however we construct it and for the first time 
the $\gamma\ne0$ version,
from first principles by a method described in appendix A. 

Using \polpsi\ and \polf, we have by the algebraic steps in \gaerg,
that \sumif\ induces a change $\delta A_t$ which acting on ${\rm
e}^{-S}$ is equivalent to \gaa.
Observing that $\partial_t\theta=0$, and computing the commutator, 
one readily discovers that
\eqn\dc{\delta A_t\equiv-\int_q 
{\rdelta\over \delta\phi_q}\left( \delta c'(q^2)
{\rdelta\over \delta\phi_{-q}}
 + 2 \left({\delta c'(q^2)\over c(q^2)}-c'(q^2)
{\delta c(q^2)\over c^2(q^2)} \right)q^2\phi_q\right)\ ,}
\ie $\delta A_t$ is just the change 
induced by $c\mapsto c +\delta c$ in \polchinski.
We have thus seen that any
infinitessimal change of the function $c$
in Polchinski's exact RG equation can be obtained through a
field redefinition.

Let us mention that a change in $\gamma(t)$ (\eg from zero to non-zero)
is also a field redefinition, whose form can be found similarly
by the methods of appendix A.\foot{{\sl despite} the fact that this maps
between exact RGs with possibly strictly different fixed point structures}\
Also note that a field redefinition induces
a reparametrization in theory space, \ie in the infinite
dimensional coupling constant space that spans all effective actions.
This latter viewpoint was pursued in ref. \sumi. Here we see that
such reparametrizations of theory space connect large classes of exact
RGs, but we stress that fundamentally all this 
follows from field redefinitions.

\newsec{Revealing the dependence in $S$}

The above discussion on field redefinitions owns its   simplicity to
 the convenient interpretation of the dependences of the kernel in the exact RG
equation $\Psi^\alpha[\phi,t]$. However, it is often the case that
we want to think of it as a function of the action $S$. 
This is, for instance, the
way we have first presented Polchinski's equation \polchinski. There 
the field functional derivatives in $\Psi[\phi,t]$ have acted on the
$\exp(-S)$ and the simple structure advocated in \wegner\ and 
\psidef\ is hidden.

To avoid 
confusion we shall denote this new functionality with a bar notation,
\eqn\newpsi{
  \Psi^\alpha[\phi,t] {\rm e}^{-S}
= \bar \Psi[\phi, S[\phi,t]] {\rm e}^{-S}\ .
}
Because the $\phi$ functional derivatives have now acted explicitly on
$S$, $\bar\Psi$ is just a function multiplying $\exp (-S)$.
The kernel dependence on $t$ is now  coming only 
implicitly through the action. (Note that $\gamma(t)$ and $S$ are not
independent: one may be regarded as a function of the other through
\polchinski.) Of course, the functional forms $\Psi$ and $\bar\Psi$ do 
transform differently under field redefinitions as we shall shortly
show. It is arguable that, though natural, phrasing the dependences
of the exact RG equation in terms of the action obscures its
transformation properties.

We are now  interested in equivalence under
transformations whose $t$ dependence is restricted in the following way
\eqn\newtheta{
 \phi \longrightarrow \tilde \phi^\alpha = \phi^\alpha-
  \bar\theta^\alpha[\phi,S]
\ .}
 Similarly, we have a new functionality for $A_t$ which we denote by
\eqn\newdeltaat{ 
 \bar A_t\equiv \bar A_t[\phi,S]\qquad ,\qquad \bar D_t\equiv \partial_t
 - \bar A_t \ .}
Upon field redefinitions, the transformation in the functional
form of $\bar A_t$ carries a new piece 
\eqn\fsda{
\delta \bar A_t[\phi,S]\equiv 
\tilde{\bar A}_t[\phi,S]-\bar A_t[\phi,S]=[\bar D_t,\bar\theta]-\delta_S
 \bar A_t\ ,
}
\ie requires subtracting the induced change $\delta_S \bar A_t$. 
This new piece amounts to the simple chain rule
\eqn\chainrule{
 \delta_S \bar A_t= \int_\psi\left(\bar\theta S[\psi,t]\right) 
{\delta \bar A_t\over \delta S[\psi,t]}
}
It is straightforward
to check that in this picture once again \simplerg\ is covariant:
since the changes are described at constant $S$, \gas\ is missing in  \gaerg,
but the new piece \chainrule\ in \fsda\ supplies it instead -- as can be seen
by using the equation that follows from operating 
$\delta/\delta S[\psi,t]$ on \simplerg. 
 
Note that at a fixed point $S=S_*$,
since we take $\bar\theta$ to depend on $t$ only through $S$, $\bar\theta$ is
also $t$ independent. This means that fixed points $S^{(1)}_*$
are mapped to fixed points $S^{(2)}_*$.
Furthermore, by the usual expansion in
first order perturbations, the image fixed point $S^{(2)}_*$ has the
same spectrum of eigenoperators as $S^{(1)}_*$. Therefore, at fixed points
$t$-independent field redefinitions have no physical effects.

Two exact RGs are equivalent under field
redefinitions if we can find a $\bar\theta$ such that we can map
from one $\bar A_t[\phi,S]$ to the other under (the exponential) of
$\bar\theta$, and this in general will reduce to differential equations for the
parts of $\bar\theta$. 
It is instructive to demonstrate \fsda\ at a less
formal level with the explicit transformation  on $\bar\Psi^\alpha$
(again with $\gamma=0$).
Substituting \psidef\ and \sumif\ into \psitranf, one finds a non-zero 
$\partial_t \bar \theta^\alpha=
(\delta c(q^2)/ 2q^2)\,{\delta \partial_t S/\delta\phi_{-q}}$, which is
expanded via \polchinski. Even after some manipulation,
the result however is not equivalent to \dc.
This is because $\delta\bar\Psi^\alpha$ in \psitranf\ does not
take into account the change induced via \density,
of $\bar \Psi^\alpha$ as a functional of $S$. From \psidef, 
one needs to subtract from \psitranf, the term 
$c'_p{\delta\over\delta\phi_{-p}}\delta S$, where from
\gas\ and \sumif,
\eqn\ds{\delta S=\int_q {\delta c(q^2)\over2q^2}
     \left( {\delta S\over \delta \phi_q}{\delta S\over \delta \phi_{-q}}
     - {\delta^2 S\over \delta \phi_q \delta \phi_{-q}}
  -2{q^2\over c(q^2)}\phi_q {\delta S\over \delta \phi_q}\right)\ .}
(Incidentally the only restriction one has on such an
infinitessimal $\delta c$ is that the
integral over $q$ in this equation is bounded, in other words that
$\delta c$ decay fast enough for large momentum as required for a 
change in cutoff function.)
Finally we find \eqnn\dpsi
$$\displaylines{
\delta\bar\Psi^\alpha=\delta c'(p^2){\delta S\over\delta\phi_{-p}}
+2\left(c'(p^2){\delta c(p^2)\over c^2(p^2)}-{\delta c'(p^2)\over c(p^2)}
\right) p^2\phi_p\hfill\dpsi\cr
\hfill
+{1\over2}\int_q\left({\delta c(p^2)\over p^2}
c'(q^2)-{\delta c(q^2)\over q^2}c'(p^2)
\right)\left({\delta^2S\over\delta\phi_q\delta\phi_{-p}}
{\delta S\over\delta\phi_{-q}}-
{\delta^3S\over\delta\phi_{-q}\delta\phi_q\delta\phi_{-p}}\right)\ ,\cr
}$$
which corresponds to the required change in \psidef, because
the second line vanishes after substitution in \wegner, by symmetry.
\bigskip

\newsec{The kernel $\Psi^\alpha$ as a connection}
We have shown that invariance
under field redefinitions allows mapping an exact
RG equation into its version in a different scheme,
as described by change of cutoff function.
We now take a higher point of view and observe that
field redefinitions further
relate different functional forms of equally valid
exact RG equations. Generically a field redefinition
induced by a $\theta^\alpha$ functional
will alter the form of the equation, and implement highly 
non-trivial changes of scheme. 

All the RG equations obtained in such
a way belong to the same `universality' class, in the sense
that no observable prediction for the continuum quantum
field theory changes.
It is apparent that exact RG equations can be transformed
under field redefinitions, which correspond to a symmetry
of the theory preserving low-energy observables. Working 
with a specific equation amounts to a choice for $\Psi^\alpha$.
The situation is reminiscent of the presence of local
symmetries. Indeed this parallelism is already evident
in our equations. 

Thus, returning to a description in terms of a direct dependence on $t$,
it is natural to interpret $D_t$ in \simplerg\ as
a covariant derivative, where from \gadf, the role of an 
antihermitian connection is played by $A_t$.
Of course \gaa\ is then nothing but a gauge transformation,
carried by $\theta$.
Note that  $\rpartial_\alpha$ is the generator of reparametrizations
(diffeomorphisms), i.e. the group we are dealing with. $A_t$ is just the 
field $\Psi^\alpha$ contracted with the generators, i.e. valued in the
Lie algebra. Similar remarks hold for $\theta$ and $\delta A_t$.  
We see from \gas, that ${\rm e}^{-S}$ transforms homogeneously
analogous to a matter field. And \gaerg\ merely confirms the now obvious
gauge covariance of \simplerg. 

Concrete
calculations at a given $t$ need  a choice of  $\Psi$, that is $A_t$,
within an orbit spanned by field redefinitions (carried out by
$\theta$). When moving to a lower cut-off, that is a different
$t$, we are allowed to choose a different set of fields
provided that a connection notifies the field redefinition
which connect the two sets. The theory is field
redefinition invariant, locally in $t$.
Geometrically, the exact RG picks a covariant section
through theory space (\aka the space of actions) fibred over RG time.

However, the gauge theory interpretation 
is far from being just a formal viewpoint.
Since this is one dimensional gauge theory, we know that
any gauge field may at least locally be transformed into
any other. In other words, locally in $t$, any exact RG
can be transformed into any other (obviously however with
the same fibres, \ie field content).
Furthermore, we can construct this transformation.
We simply consider a Wilson line starting from say $t=t_0$:
\eqn\Ph{\Phi(t)=P\exp\left(-\int^t_{t_0}\!\!\!\!ds\, A_s\right),}
where $P$ is path ordering, placing as usual the later $A$s to the right. 
Thus $\partial_t\Phi(t)=-\Phi(t) A_t$ and $\partial_t\Phi^{-1}(t)
=-A_t\Phi^{-1}(t)$.
Therefore we have immediately,
\eqn\triva{A_t=-\Phi^{-1}(t)\,\partial_t\,\Phi(t)\ .}
By \gaa, $\delta D_t=-[D_t,\theta]$, and thus \triva\ 
is a finite gauge transformation 
of $A_t=0$. Since every
gauge field is thus equivalent to $A_t=0$, over any finite range in $t$,
every gauge field is equivalent
to each other over any finite range.
Indeed by substituting \triva\ into \simplerg, we obtain that any exact RG
is equivalent to the trivial $\Psi^\alpha=0$ equation, 
\eqn\triveq{\partial_t\, {\rm e}^{-S_0}=0\ ,}
where
\eqn\triv{{\rm e}^{-S_0}=\Phi(t)\,{\rm e}^{-S}\ .}
By \Ph, $S_0$ is nothing but $S$ at $t=t_0$.
Inverting this relation, we thus can write the solution of any exact
RG in closed form as a series of integrals:
\eqn\trivii{{\rm e}^{-S}=\Phi^{-1}(t)\,{\rm e}^{-S_0}\ .}
It is easy to
convince oneself that the momentum integrals obtained from expanding 
the notation in \trivii\ converge, 
the result thus being a well defined series
for $\exp(- S)$. 

Let us stress that the conclusion arrived at above,
that every exact RG, \ie $A_t$, is equivalent to every other
over any finite range of $t$, holds when $A_t$ is expressed  
as a function of $\phi$ and $t$ only. Changing variables to 
$\phi$ and $S$ in order to properly 
compare exact RGs, as in \newpsi\ and sec. 4, means that
such an equivalence may then hold only over a limited domain or 
indeed break down entirely
(for example at a fixed point), because of the difficulty of inverting
this change of variables. There are also global obstructions
to equivalence that can be understood from the gauge theory interpretation.

So far, we have been discussing non-observable quantities. A gauge
field can be made to vanish at a particular point, but gauge-invariant
observables
retain its traces, and these are also obstructions
to completely gauge away $A_t$. Generally, these obstructions
can be either local or non-local, but since the
RG equation describes a one-dimensional flow in $t$, there
are no local invariants which can be build from $A_t$. For instance, the
equivalent of the field strength is identically zero in spite of the
fact that the group of field redefinitions is non-Abelian. We are bound
to use non-local invariants. Given that there are no flows that form
closed loops in unitary Poincar\'e invariant theories\foot{corresponding
to real Euclidean invariant theories, in Euclidean
space.} \ref\josec{S. Forte and J. I. Latorre, Nucl. Phys. B535 (1998) 709-728;
    hep-th/9805015.},
the only remaining candidate is
\eqn\Ph{\Phi=P\exp\left(-\int_{-\infty}^\infty 
 \!\!\!\! ds\, A_s\right),
}
providing these limits exist. However, the existence of these limits
implies that $A_t\to0$ there, \ie that we consider solutions of
\wegner\ that begin ($t=-\infty$) and end ($t=+\infty$) at fixed points.
This quantity of course transforms homogeneously 
at both extremes of the flow with $t$-independent field redefinitions. To be
more precise, the path-ordered line that connects the
initial ($i$) fixed point with the final ($f$) one does transform 
under general field redefinitions as
\eqn\bilocal{
 \Phi_{fi} \longrightarrow \Omega_f \Phi_{fi} \Omega_i^{-1}\ ,}
where $\Omega_{i,f}$ are $t$-independent finite field redefinitions at
the initial and final fixed points, respectively. Thus we have
\eqn\flow{\exp -S^f_* = \Phi_{fi} \exp -S^i_* \ .
}
Given that the two theories display different universality properties,
it is then impossible to gauge away $A_t$ since $\Phi^{fi} $ would then 
be unity modulo end-point field redefinitions. 
Equally since such a relation \flow\ transforms covariantly
under field redefinitions, two exact RGs which are equivalent under
field redefinitions must have the same network of flows between
fixed points displaying the same spectrum of eigenoperators.
And on the contrary, for two exact RGs that do not display equivalent 
fixed points with the same spectrum and network of flows, there can
be no exponentiated solution $\theta$ mapping one to the other via
\fsda. 

We can rephrase the above argument in more physical terms. The inherent
freedom to use field redefinitions allows for deforming at will the
pace of integration of modes along the RG trajectory. It is always possible
to completely stop the integration of modes along a finite
part of the path. Nevertheless, the above obstruction to gauge away 
the connection means that modes will have to be integrated somewhere else. 
The extreme (unphysical) case would correspond to integrate all modes at a 
single scale. Looking back at Wilson's original deduction, it is clear
that such freedom was present in the choice of scheme defined through his
$\alpha$ function \kogwil. Our analysis is, though, more general than
such a simple scheme change.

Let us end up this section making clear the idea that
field redefinitions play a double r\^ole in exact RG equations:\par
$\bullet$ The kernel $\Psi^\alpha$ of an exact RG equation is shaped as
a field redefinition.\par
$\bullet$ A change of this kernel, i.e. of scheme, is implemented via a field
redefinition.
\par\noindent The latter effect of  field redefinitions is natural
and shows intuitively that scheme changes are immaterial to
 physical observables. Instead, the first r\^ole may be
 confusing as it seems to  go against the (irreversible) 
integration of modes picture underlying the RG. 
This paradox can be traced to the fact that $A_t$ can be locally gauged
away due to the 1-dimensional (that is in $t$)  nature of the flow.

This element of confusion would be absent in other contexts, e.g.
a standard 4-dimensional gauge theory where $A_\mu$ cannot be gauged away.
There, $A_\mu$ carries  physical
  particles  as it is further reflected in the 
obstruction to get rid of the field strength $F_{\mu\nu}$.
In the case of exact RG equations the very nature of the equation is 
attached to its kernel $\Psi^\alpha$ which cannot be gauged away globally
as quantified by the $t$-ordered Wilson line $\Phi$.

\newsec{Conclusions}

Exact renormalization group equations make quantitative the statement that 
low-energy physics is independent of the details of the cut-off. A specific
 change of that cut-off can be absorbed into a suitable 
field redefinition. This idea leads hierarchically  
to the standard Callan-Symanzik covariance statement in perturbation theory.

A particular exact RG equation is characterized by its kernel $\Psi^\alpha$.
The freedom to choose the form of this kernel is thus  related to field
redefinitions. Infinitessimally, two kernels, expressed in terms of the
action $S$, are physically equivalent
 if they are related by \fsda. Globally, two kernels descibe the same
RG flow between fixed points $i$ and  $f$ if their
 corresponding connections $A_1$ and $A_2$ produce
$t$-ordered lines which are related by
\eqn\finaleq{
\Phi_{fi}[A_1]= \Omega_f \Phi_{fi}[A_2]\Omega_i^{-1}\ ,}
where $\Omega_{i,f}$ are $t$-independent field redefinitions of 
the fixed points actions at points $i$ and $f$. 
In perturbation theory, an operator basis in chosen and all field
redefinitions turn into finite coupling constant redefinitions. 
A change of perturbative scheme, that is of local counterterms,
is tantamount to a finite redefinition of coupling constants.

Of course field redefinitions could be chosen which obscure the symmetries
of the field theory. Equally, symmetries that
are apparently broken by some exact RG, through
its cutoff function or/and other parts of its structure, may only
be deformed, rather than lost. This would be established if (the exponential
of) a $\bar\theta$ satisfying \fsda\ could be found that maps the exact RG
to one preserving the symmetry.

Freedom under field redefinition should be used for
the practitioner's benefit. Though physics is independent of
such redefinitions it is obvious that practical schemes 
do truncate in one way or another such freedom. Within a particular
truncation, some {\sl a priori} equivalent forms of
exact RG equations will differ. Accurate results for observables
may be approached faster in a particular renormalization scheme. 
Using the above results it is possible to construct 
a RG equation which carries higher functional derivatives
than Polchinski's. Such an equation could be contrived
so that its local potential approximation \ref\lpa{J.F. Nicoll,
T.S. Chang and H.E. Stanley, Phys. Rev. Lett. 33 (1974) 540.}\
is two-loop exact. Truncations, in particular this and
higher orders in the derivative expansion, may well be more accurate
than the corresponding results from the standard flow equations.

\bigskip
\goodbreak
\noindent{\bf Acknowledgments}
\nobreak
J.I.L. acknowledges financial support from 
CICYT (contract AEN98-0431) and
CIRIT (contract 1998SGR-00026). T.R.M. acknowledges financial
support of PPARC grant GR/K55738. We both acknowledge
the financial support of a BC-MEC Acciones Integradas
grant MDR(A998/99)1799.

\vfill\eject

\appendix{A}{Finding the field redefinition corresponding to changing cutoff}
If we restore a non-zero $\gamma(t)$, $A_t$ may still be written equivalently
as \polpsi. We require to find $\theta[\phi,t]$ such that formula \gaa\
reproduces \dc. It is helpful to introduce the operators:
\eqn\ops{\eqalign{
\Delta^0[a]&=\int_p a(p^2) 
{\rdelta\over \delta\phi_{p}}\phi_p\ ,\qquad\cr
\Delta^1&=\int_p  
{\rdelta\over \delta\phi_{p}}p^\mu{\partial\over p^\mu}\phi_p\ ,\qquad\cr
\Delta^2[a]&=\int_p a(p^2)
{\rdelta\over \delta\phi_{p}}{\rdelta\over \delta\phi_{-p}}\ .\cr
}}
They form a closed algebra:
\eqn\algops{\eqalign{
\left[\Delta^0[a],\Delta^0[b]\right]&=\left[\Delta^2[a],\Delta^2[b]
\right]=0\cr
\left[\Delta^0[a],\Delta^2[b]\right]&=-2\Delta[ab]\cr
\left[\Delta^1,\Delta^0[a]\right]&=-2\Delta^0[p^2a']\cr
\left[\Delta^1,\Delta^2[a]\right]&=\Delta^2[Da-2p^2a']\ .\cr
}}
We may readily express \polpsi\ and \dc\ in terms of these. Taking as an
ansatz 
\eqn\ansath{\theta=\Delta^0[H(p^2,t)]+\Delta^2[K(p^2,t)]\ ,}
we find from \gaa\ and \algops:
\eqn\ansode{\eqalign{
{\partial\over\partial t}H&+2p^2H'=-2p^2\left(\delta c/c\right)'\cr
{\rm and}\qquad
{\partial\over\partial t}K&+(2-\gamma)K+2p^2K'+2Hc'-4p^2Kc'/c=-\delta c'\ .\cr
}}
[Here $c\equiv c(p^2)$.] We take the simplest solution to the first equation:
$H=-\delta c/c$, after which the second can be solved by the method of
characteristics:
$$
K(p^2,t)=-{\delta c(p^2)\over2p^2}
-{c^2(p^2)\over2p^2}\ {\rm e}^{\Gamma(t)}\!\int^t\!\!\!\!ds\ 
{\rm e}^{-\Gamma(s)}\ \gamma(s)\,
{\delta c(p^2\,{\rm e}^{2s-2t})\over c^2(p^2\,{\rm e}^{2s-2t})}\ ,$$
where $\Gamma(t) =\int^t\!ds\ \gamma(s)$. In the case that $\gamma=0$,
this solution collapses to \polf.

\listrefs
\bye